\documentclass[journal=jacsat,manuscript=article]{achemso}

\usepackage[version=3]{mhchem} 
\usepackage[T1]{fontenc}       
\usepackage{graphicx}

\newcommand{\mm}[1]     {\ifmmode {#1} \else{}${#1}$\fi}
\newcommand{\mmm}[1]    {\ifmmode{}#1 \else{}${#1}$\fi}
\newcommand{\beq}[1]    {\begin{equation} \label{#1} }
\newcommand{\eeq}       {\end{equation}}

\def \tmo{\mm{\rm Tm_{2}Mn_{2}O_7}}

\def \ot{{\mm{{1\over2}}}}
\def \oe{{\mm{{1\over8}}}}
\def \te{{\mm{{3\over8}}}}
\def \of{{\mm{{1\over4}}}}
\def \tf{{\mm{{3\over4}}}}

\author{E.~Pomjakushina}
\email{ekaterina.pomjakushina@psi.ch}
\affiliation{Laboratory for Developments and Methods, PSI, 5232
Villigen, Switzerland}
\phone{+41 (0)56 310 3207}
\fax{+41 (0)56 310 2939}
\author{V.~Pomjakushin}
\affiliation{Laboratory for Neutron Scattering and Imaging,PSI, 5232
Villigen, Switzerland}
\author{K.~Rolfs}
\affiliation{Laboratory for Developments and Methods, PSI, 5232
Villigen, Switzerland}
\author{J.~Karpinski}
\affiliation{Laboratory for Developments and Methods, PSI, 5232
Villigen, Switzerland}
\author{K.~Conder}
\affiliation{Laboratory for Developments and Methods, PSI, 5232
Villigen, Switzerland}

\title{\large New synthesis route and magnetic structure of $\rm Tm_{2}Mn_{2}O_{7}$ - pyrochlore}

\abbreviations{FM,AFM,TG}

\begin{document}
\def\figsiz{\columnwidth}
\def\figsiz{\textwidth}
\def\figsiz{15cm}

\begin{tocentry}
A pure phase $\rm Tm_{2}Mn_{2}O_{7}$-pyrochlore compound (space group $Fd\bar{3}m$) was obtained by converting a hexagonal $\rm TmMnO_{3}$ (space group $P6_{3}cm$) at 1100 $^o$C and 1300 bar oxygen pressure.
This is a new promising route to stabilize Mn$\rm^{4+}$ pyrochlores  thermodynamically unstable at ambient pressure. The unique high temperature high pressure apparatus used in this work allows to synthesize materials in cm$^3$ amount, sufficient to collect good quality neutron scattering data.
\end{tocentry}

\begin{abstract}

 In this work we present a new chemical route to synthesize $\rm Tm_{2}Mn_{2}O_{7}$ pyrochlore, a compound which is thermodynamically unstable at ambient pressure.  Differently from the reported in the past high pressure synthesis of the same compound applying oxides as starting materials, we have obtained a pure phase $\rm Tm_{2}Mn_{2}O_{7}$  by a converting $\rm TmMnO_{3}$ at 1100 $^\circ$C and 1300 bar oxygen pressure.  The studies of $\rm Tm_{2}Mn_{2}O_{7}$ performed by a high resolution
neutron powder diffraction  have shown, that  a pure pyrochlore cubic phase $\rm Tm_{2}Mn_{2}O_{7}$ (space group $Fd\bar{3}m$) have been obtained. On cooling below 25K there is a transition to a ferromagnetically (FM) ordered phase observed with an additional antiferromagnetic (AFM) canting suggesting a lowering of the initial cubic crystal symmetry. The magnetic transition is accompanied by small but well visible magnetostrsiction effect. Using symmetry analysis we have found a solution for the AFM structure in the maximal Shubnikov subgroup $I4_1/am'd'$.

\end{abstract}


\maketitle


\section{Introduction}
A few important factors influence any successful synthesis of  novel electronic materials: a reactivity
of the used precursors, an applied type of the synthesis process (solid state reaction, reactions in a liquid phase, reaction in a gas phase),
and the thermodynamic parameters. Beside temperature also pressure is a fundamental thermodynamic variable of every chemical reaction. An influence of the pressure on a solid state reaction depends on the kind of the applied pressure medium. With applying just a hydrostatic pressure, a reduced evaporation of the volatile components during synthesis, a structural rebuilding of the resulting solid  (bond lengths and coordination number change) and higher reactivity (reacting powder grains are brought closer together) can be expected. If the applied gas medium is an active component of the system (such as oxygen in the case of oxides) the thermodynamic equilibrium will be changed. This causes modification of the phase diagrams and changes in the defect equilibria of a solid. In the case of oxidizing gases, especially oxygen, this  can result in a stabilization of the unusually high oxidation states of cations and allow synthesis of compounds thermodynamically unstable in air, that is at low oxygen partial pressure.

Pyrochlore oxides containing transition metals show geometrical magnetic frustration, which is interesting because of a possible realization of novel, exotic, magnetic ground states. The cubic pyrochlore oxides ($Fd\bar{3}m$) have a general formula $\rm A_{2}B_{2}O_{7}$, where A is usually a trivalent rare earth metal and B is a tetravalent transition metal. Pyrochlore structure is based on a network of corner-sharing $\rm BO_{6}$ octahedra similar as observed in perovskites $\rm ABO_3$. The A atoms are in the center of the hexagonal rings formed by oxygens (O1, located at the 48f positions) with two more oxygens (O2, at the 8b positions) located above and below the rings.  Both A and B cations in this structure are located in the corners of tetrahedra. Such configuration can lead to a geometrical magnetic frustration.

There are almost 20 tetravalent ions (B-site), which can form pyrochlore compounds with rare earth ions (A-site). Subramanian and Sleight \cite{Subramanian1993225} have built a stability-field map for these materials, which displays the stability limits for the pyrochlore structure at ambient pressure depending on a  so called stability range factor defined as ionic radius ratio $\rm r_{A^{3+}}/r_{B^{4+}}$. For stability at ambient pressure it lays in the range between 1.36 and 1.71. Based on this map it looks that the rare earth Mn$\rm^{4+}$ pyrochlores can not be synthesized at ambient pressure and require high pressures for the stabilization. This is mainly due to the small size of Mn$\rm^{4+}$ ion compared to the trivalent rare earth cations  giving the stability range factor $>$ 1.84.

The first report about successful synthesis of $\rm A_2Mn_2O_7$ (A= Y, Tl) appeared in 1979 by Fujinaka et al. \cite{Fujinaka1979}. Both compounds were synthesized at 1000-1100$^\circ$C and pressures 3-6GPa.
About ten years later  high pressure synthesis of $\rm A_2Mn_2O_7$ (A=Sc, Y, In, Tl, Tb-Lu) at pressures 5-8GPa and temperatures 1000-1500$^\circ$C \cite{Troyanchuk1988} and A=Y, Dy-Lu by a hydrothermal method in sealed gold ampules at 0.3 GPa and 500$^\circ$C  \cite{Subramanian1988} was reported. Later Shimakawa et al. \cite{Shimakava1999}  used a hot isostatic press method at
1000-1300$^\circ$C and only 0.4 kbar for the series with A=In, Y, and Lu while for A=Tl material still 2.5 GPa and 1000$^\circ$C in a piston-cylinder apparatus was required.
All the above synthesises were  done in anvils with solid state pressure medium without controlled oxygen pressure atmosphere.
According to the recent review of Gardner et al. \cite{Gardner_RevModPhys_2010} all these materials appear to be ferromagnets. Although a variety of the preparative conditions were applied, an excellent agreement  of the structural parameters was achieved by different groups is excellent.

In comparison to other pyrochlore families such as $\rm A_{2}Mo_{2}O_{7}$, $\rm A_{2}Ti_{2}O_{7}$, $\rm A_{2}Sn_{2}O_{7}$, $\rm A_{2}Ru_{2}O_{7}$ \cite{Gardner_RevModPhys_2010}, $\rm A_{2}Mn_{2}O_{7}$ system was not  much explored (especially some compositions as $\rm Tm_{2}Mn_{2}O_{7}$ for example), perhaps due to ambient pressure instability and consequently a non trivial synthesis.
To the best of our knowledge there were only two groups at the end of eighties reporting on the synthesis of $\rm Tm_{2}Mn_{2}O_{7}$ \cite{Troyanchuk1988,Subramanian1988}.
Magnetisation studies were done for $\rm Tm_{2}Mn_{2}O_{7}$. Discrepant values for the Curie temperature $T_C$ (14K \cite{Troyanchuk1988} and 30(5)K \cite{Subramanian1988}) and the Curie-Weiss constant $\Theta$ (+56(8)) \cite{Subramanian1988}) were reported. The observed large and positive $\Theta$ value indicates strong ferromagnetic exchange interactions between the Mn moments. At low temperatures a sharp increase in the susceptibility again indicating a possibility for a long-range ferromagnetic order was observed.

In this work we report new studies of the magnetic properties of $\rm Tm_2Mn_2O_7$  obtained by an oxidation of $\rm TmMnO_3$ at 1100 $^\circ$C and 1300 bar oxygen pressure.

As $\rm Tm^{3+}$ cation has one of the smallest radius among the rare earth elements and consequently the tolerance factor is also the smallest, it  has a big magnetic moment and a low neutron absorbtion and no neutron diffraction studies were performed for $\rm Tm_{2}Mn_{2}O_{7}$ up to now, we have decided to undertake investigation of this compound, synthesized by a novel high pressure oxygen method.

\section{Experimental section}
\subsection{Materials and synthesis}
The following reagents were used: $\rm Tm_{2}O_{3}$ (99.99\% Aldrich) and $\rm MnO_{2}$ (99.997\% Alfa Aestar). At first  $\rm TmMnO_{3}$ was synthesized by a standard solid state reaction. The respective amounts of the starting reagents were mixed, milled and
calcined at 1000 - 1200$\rm ^{o}$ C during 100 h in air with intermediate grindings.
The synthesized material was found to be  phase pure (hexagonal phase) as proved
by laboratory powder X-ray diffraction.  $\rm Tm_{2}Mn_{2}O_{7}$ was obtained by an oxidation of 3 g of the starting material at 1100$\rm ^{o}$ C and 1300 bar oxygen pressure during 20 hours in the high pressure setup \cite{Karpinski}. The volume of oxygen used during the synthesis in in this setup
is reduced to 20cm$^3$ by using double-chamber construction.
The sample was placed in an alumina crucible connected tightly to an oxygen supply system of the autoclave. The crucible was placed in the furnace in the high-pressure chamber. In this chamber the oxygen pressure in the crucible is balanced by the equal argon pressure outside the crucible.  In this
way, the volume of the oxygen under high pressure is only  up to 20cm$^3$, whereas the volume of the whole chamber is about 4000cm$^3$. Both pressures were balanced by an electronic pressure control unit with remote control valves.
\subsection{Measurements and characterisation}
Oxygen content in the sample was determined by a thermogravimetric (TG) hydrogen reduction  performed on NETZSCH STA 449C analyser equipped with PFEIFFER VACUUM ThermoStar mass spectrometer. The same TG analyser was used for the determination of the temperature stability range of the sample in pure helium and artificial air (21 \%(vol.) oxygen in helium).
AC and DC magnetization ($M_{AC}/M_{DC}$) measurements were performed using Quantum Design PPMS  at
temperatures ranging from 2 to 300~K. The AC field amplitude and the
frequency were 0.1~mT and 0.05 ~mT and 1000~Hz, respectively.
Neutron powder diffraction measurements were carried out at the
high-resolution HRPT diffractometer \cite{hrpt} at SINQ neutron
spallation source (PSI, Switzerland).   The refinements of the
crystal structure parameters were done using {\tt
FULLPROF}~\cite{Fullprof} program, with the use of its internal
tables for neutron scattering lengths.
The symmetry analysis was performed using {\tt ISODISTORT} tool based on {\tt ISOTROPY} software \cite{isot,isod}, {\tt BasiRep} program \cite{Fullprof} and software tools of Bilbao crystallographic server \cite{Bilbao}.

\section{Results and discussion}

\subsection{Phase purity, oxygen stoichiometry, thermal stability}
\begin{figure}[p]
\begin{center}
\leavevmode
\includegraphics[width=\figsiz]{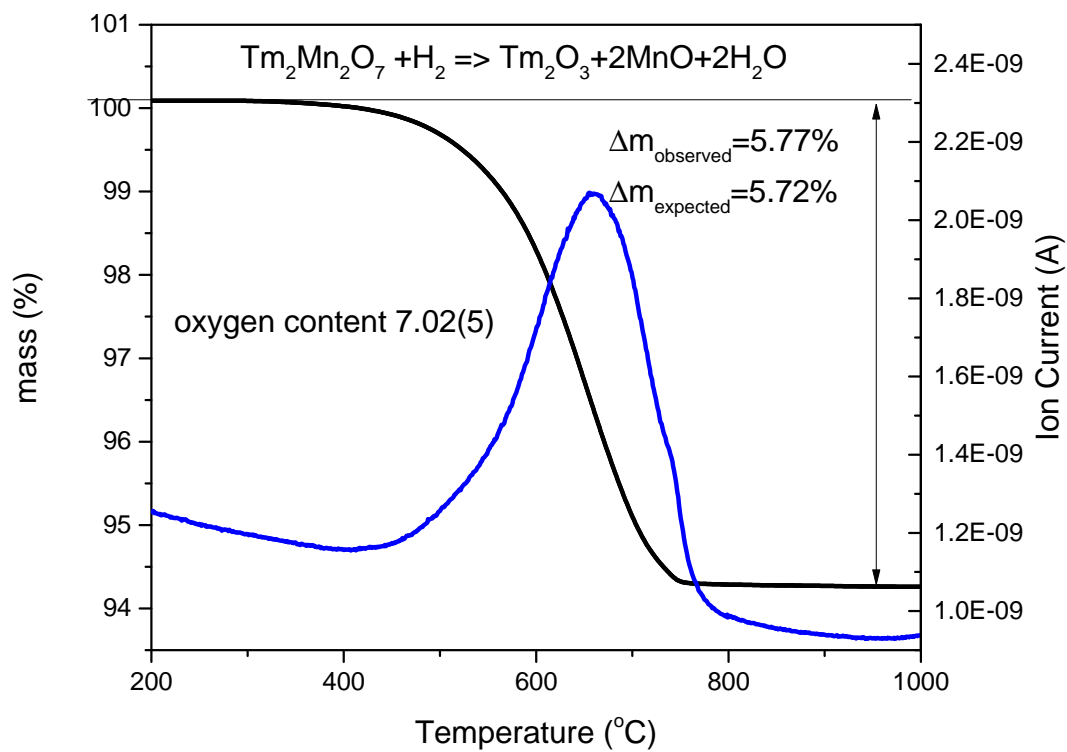}
\end{center}
\caption{Thermogravimetric curve of the hydrogen reduction of $\rm Tm_{2}Mn_{2}O_{7}$ sample. The blue line shows a mass spectrometric signal for water (m/e = 18).} \label{TG_H2}
\end{figure}

\begin{figure}[p]
\begin{center}
\leavevmode
\includegraphics[width=\figsiz]{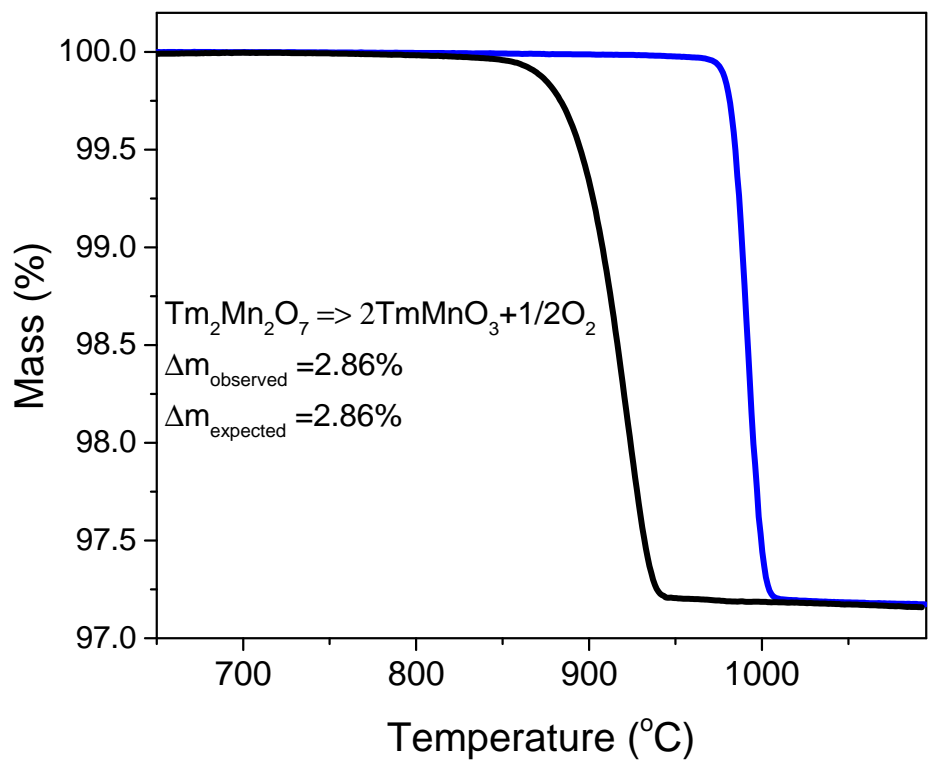}
\end{center}
\caption{Thermogravimetric curves of $\rm Tm_{2}Mn_{2}O_{7}$ sample heated in helium (black line) and artificial air (blue line).} \label{TG2}
\end{figure}

 The laboratory X-ray diffraction measurements,  which were done at room temperature using Cu K$_\alpha$ radiation  on a Brucker D8 diffractometer, have proven that the the product of the oxidation reaction is a phase pure compound with the cubic pyrochlore  structure. Oxygen content  determined by the  thermogravimetric hydrogen reduction \cite{Conder_MatResBul2005} was found to be 7.02(5).  Figure~\ref{TG_H2} shows a thermogravimetric curve of the hydrogen reduction of $\rm Tm_{2}Mn_{2}O_{7}$ sample together with a mass spectrometric signal for the water created during reduction (m/e = 18).
The thermal stability range of the synthesized compound was also checked by thermogravimerty. Figure~\ref{TG2} shows thermogravimetric curves of $\rm Tm_{2}Mn_{2}O_{7}$ sample heated in helium  and artificial air. The thermal stability of the compound increases with the increase of the oxygen partial pressure. In helium $\rm Tm_{2}Mn_{2}O_{7}$ starts to decompose at 850$^\circ$, while in air it is stable up to 950$^\circ$. Using laboratory x-ray we have identified the product of the thermal decomposition, which is hexagonal $\rm TmMnO_{3}$.
Opposite to the previous reports, where the synthesis was done starting from  oxides\cite{Troyanchuk1988,Subramanian1988}, here we have obtained a pure phase compound by converting $\rm TmMnO_{3}$ at 1100 $^\circ$C and 1300 bar oxygen pressure into $\rm Tm_{2}Mn_{2}O_{7}$.
This is a new promising route to stabilize Mn$\rm^{4+}$ pyrochlores  thermodinamically unstable at ambient pressure. The chemical reaction is reversible, and on heating starting from 850$^\circ$ (at low oxygen partial pressures) $\rm Tm_{2}Mn_{2}O_{7}$ transforms back to $\rm TmMnO_{3}$.

\subsection{Magnetic properties}
\begin{figure}[p]
\begin{center}
\leavevmode
\includegraphics[width=\figsiz]{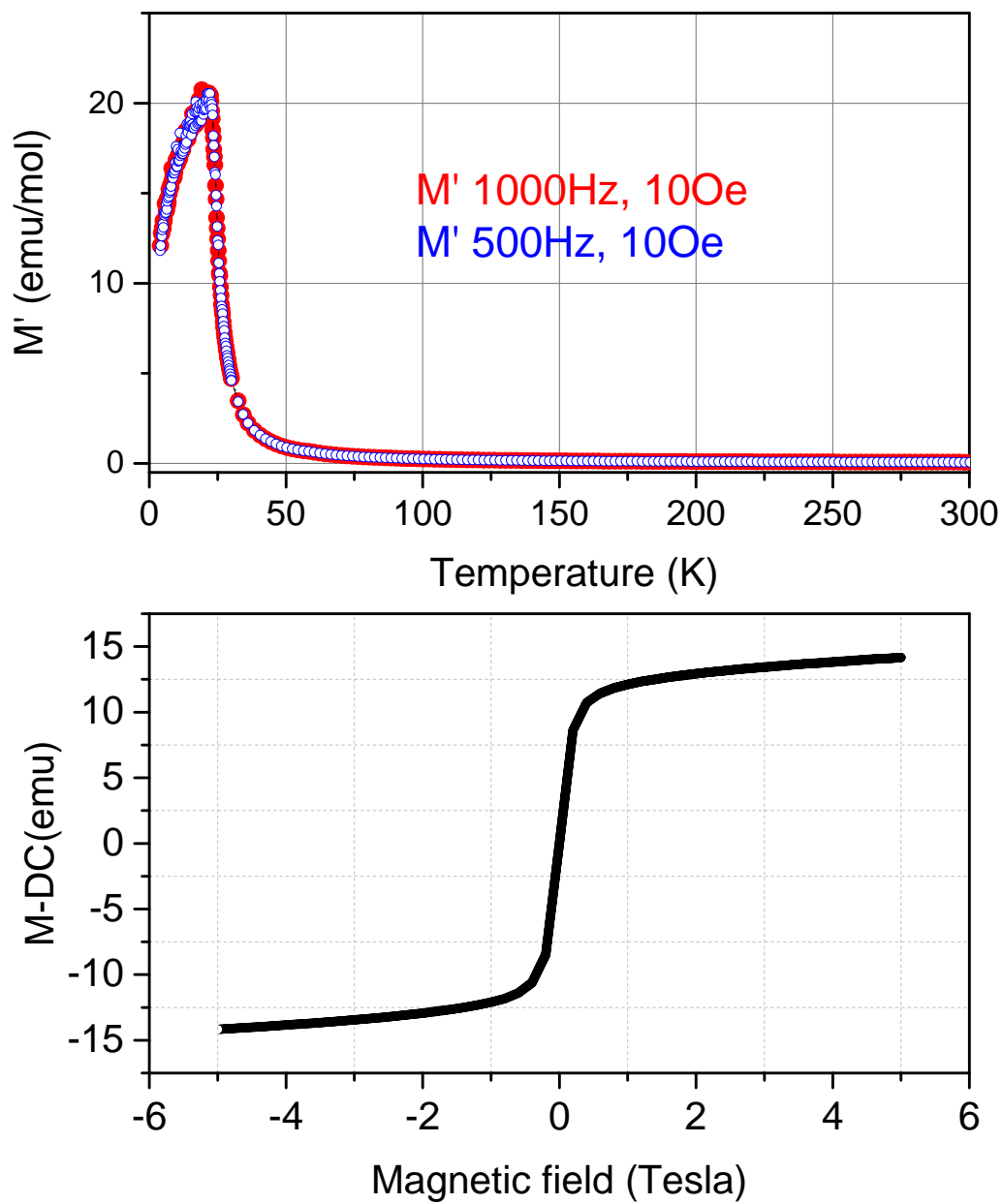}
\end{center}
\caption{(a). The temperature dependence of the real part of the AC susceptibility measured at 0.1~mT (blue) and 0.05 ~mT (red) and 1000~Hz. (b). The magnetic moment as a function of applied field measured at 5K.} \label{magn}
\end{figure}

Figure~\ref{magn}(a) shows the real part of the AC susceptibility measured at 0.1~mT and 0.05 ~mT and 1000~Hz indicating a very sharp  maximum
at temperature around 25K, similar to the observed in  $\rm Y_{2}Mn_{2}O_{7}$ \cite{ReimersGreedan_PRB1991}, $\rm Ho_{2}Mn_{2}O_{7}$ and $\rm Yb_{2}Mn_{2}O_{7}$ \cite{Greedan_Raju_PRB1996}. On the Figure~\ref{magn}(b) magnetic moment as a function of the applied field is plotted for $\rm Tm_{2}Mn_{2}O_{7}$ at 5K that is below the magnetic transition. No saturation magnetization behavior was observed up to 5 T. This
was already previously noticed for $\rm Tm_{2}Mn_{2}O_{7}$ by Troyanchuk at al \cite{Troyanchuk1988}. The authors \cite{Troyanchuk1988} proposed that the value of the f-d exchange in this compound is smaller than that in other rare earth-manganese pyrochlores and the magnetic moment of the Tm$\rm^{3+}$ ion is in a paramagnetic state.

\subsection{Magnetic structure, symmetry analysis}
\begin{figure}
\begin{center}
\includegraphics[width=\figsiz]{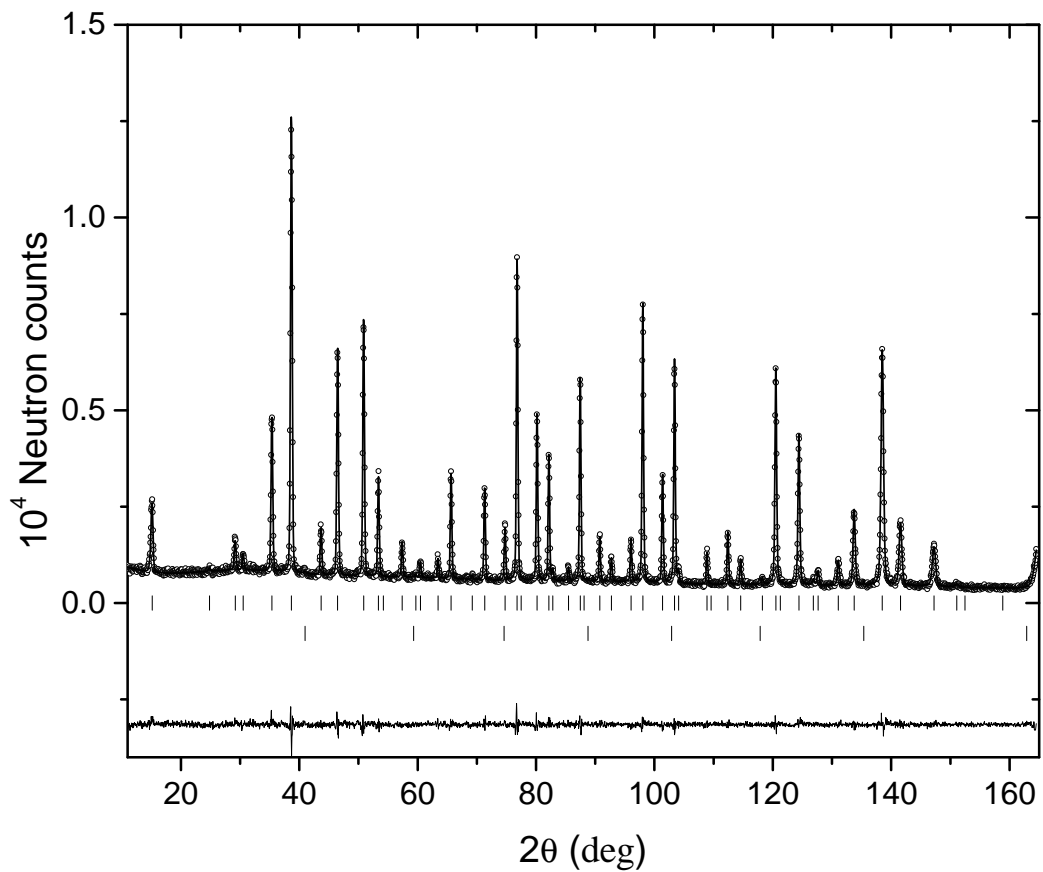} 
\end{center}

\caption{The Rietveld refinement pattern and the difference plot of the neutron diffraction data for the sample \tmo\ at T=30K measured at HRPT with the wavelength $\lambda=1.494$~\AA. The rows of tics show the Bragg peak positions for the main phase and for V container.}
\label{diff30k}
\end{figure}

The neutron diffraction studies have shown, that the synthesized material contains a pure phase of a well known  \cite{Subra1997,Gardner_RevModPhys_2010} pyrochlore cubic structure with the space group $Fd\bar{3}m$ (no. 227). Figure~\ref{diff30k} shows the powder neutron diffraction pattern and the Rietveld refinement curve for \tmo\ at T=30K, above the Curie temperature $T_C=25$~K. Structural parameters are listed in the Table~\ref{tab_str}. Note that in the pyrochlore structure Tm and Mn occupy the positions in the inversion centres and their positions can be swapped, provided that the oxygen positions are also accordingly modified. So, in this second equivalent description of the structure oxygen O1 occupies 8b-position (\te,\te,\te), O2 remains at the same (48f)-position, but $x$ should be changed to -$x$+\tf.

\begin{figure}
\begin{center}
\includegraphics[width=\figsiz]{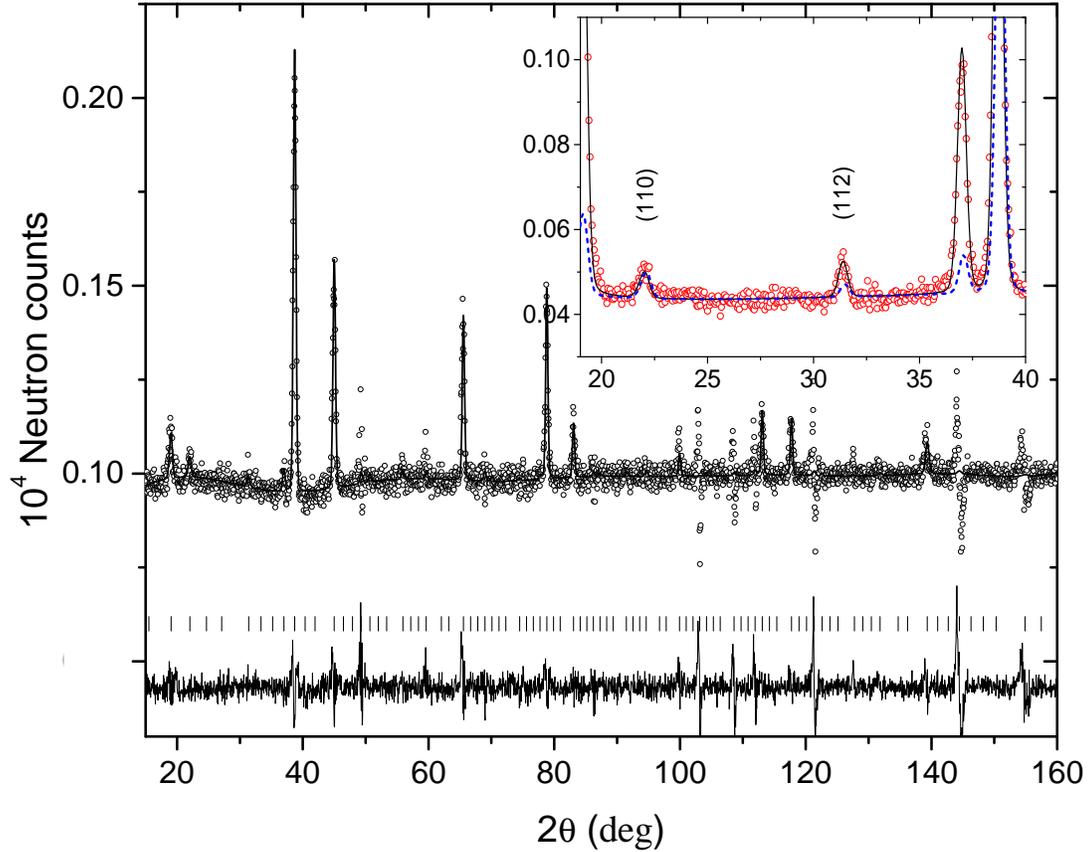} 
\end{center}

\caption{The Rietveld refinement pattern and the difference plot of the
    difference magnetic neutron diffraction pattern between 5K and
    30K  for \tmo\ measured at HRPT with the wavelength $\lambda=1.886$~\AA.
The line is the refinement patter based on the magnetic model $I4_1/am'd'$ shown in Table~ \ref{tab_str}. The rows of tics show the magnetic Bragg peak positions. The inset shows fragment of diffraction pattern at T=5K with the dashed line showing the magnetic contribution. The indexing is given in the Shubnikov group [in the parent paramagnetic group the indexes are (200) and (220)].  The indicated Bragg peaks have magnetic contribution solely from AFM components of the spins.}
\label{diff5-30k}
\end{figure}

Figure~\ref{diff5-30k} shows the Rietveld refinement pattern and the difference plot of the difference magnetic neutron diffraction pattern between 5K and 30K. All magnetic peaks are indexed with propagation vector k=0. The main contribution to the magnetic intensity originates from the ferromagnetic ordering of both types of magnetic atoms Mn and Tm. However, there is a small but well visible AFM-contribution. For example, the magnetic Bragg peaks (200) and (220) shown in the inset of \ref{diff5-30k} have no FM contribution. The experimental intensity of the very first magnetic peak  (111) is also underestimated without AFM contribution.
The magnetic atoms Mn and Tm occupy the positions (16d) and (16c), which have the same symmetry $.\bar{3}m$. The decomposition of the magnetic representation for them into irreducible representations (irreps) of $\Gamma$-point reads $\Gamma_2^{+}(\tau_3) \oplus \Gamma_3^{+}(\tau_5) \oplus \Gamma_5^{+}(\tau_7) \oplus 2\Gamma_4^{+}(\tau_9)$ with the dimensions of irreps 1D, 2D, 3D and 3D respectively. The nomenclature for the irreps is given according to Ref.~\cite{isod} with Kovalev's notation in the parentheses.
In is known from the magnetic susceptibility data and also from the previous neutron diffraction works \cite{Gardner_RevModPhys_2010} on similar pyrochlores, that the magnetic structure is ferromagnetic (FM). This fact imposes restrictions on the possible irrep involved in the magnetic transition. One dimensional real irrep $\Gamma_2^{+}$ resulting in $Fd\bar{3}m'$ Shubnikov magnetic space group (MSG) does not allow FM configuration. Two dimensional real irrep $\Gamma_3^{+}$  does not allow FM, as well for any direction of the order parameter (OP) in the configuration space, resulting in magnetic group $Fddd$ for the general OP-direction $(ab)$. The magnetic modes generated by three dimensional real irrep $\Gamma_5^{+}$ are also non-FM. The only irrep that allows ferromagnetism is $\Gamma_4^{+}$. However, the representational approach  alone is not of the big help here because this 3D irrep enters two times in the magnetic decomposition resulting in at least six independent normal modes for each sort of the atoms, i.e. twelve parameters to be refined. This general case of irrep $\Gamma_4^{+}$ corresponds to OP direction $(abc)$ and lowers symmetry down to the triclinic Shubnikov group $P\bar{1}$. More restrictive by symmetry approach is to use Shubnikov magnetic group symmetry. There are three special directions of the order parameter $(a00)$, $(aa0)$ and $(aaa)$ that generate $I4_1/am'd'$, $Imm'a'$ and $R\bar{3}m'$  isotropy subgroups, respectively. Each of the above OP directions result in only one magnetic mode for $\Gamma_4^{+}$ (magnetic mode defines specific magnetic configuration on all atoms of the same type and has only one parameter to be determined experimentally - it's amplitude). The first two subgroups are the maximal subgroups of the grey paramagnetic group $Fd\bar{3}m1'$. The group $R\bar{3}m'$ is not the maximal subgroup of the paramagnetic group, but is a subgroup of non-FM maximal subgroup $Fd\bar{3}m'$, which is generated by $\Gamma_2^{+}$. For this reason, there is an additional independent antiferromagnetic mode in  $R\bar{3}m'$ that makes magnetic atoms of the same kind inequivalent. The magnetic group $Imm'a'$ is a maximal subgroup and generated by a single irrep, but accidentally it can be also generated by $\Gamma_5^{+}$ with OP direction $(0a$-$a)$. This is a rare exceptional case when two different irreps have the same MSG as the isotropy subgroup. So, for the magnetic group $Imm'a'$ one has an additional AFM mode as well, similar to the case of $R\bar{3}m'$.

\begin{figure}
\begin{center}
\includegraphics[width=\figsiz]{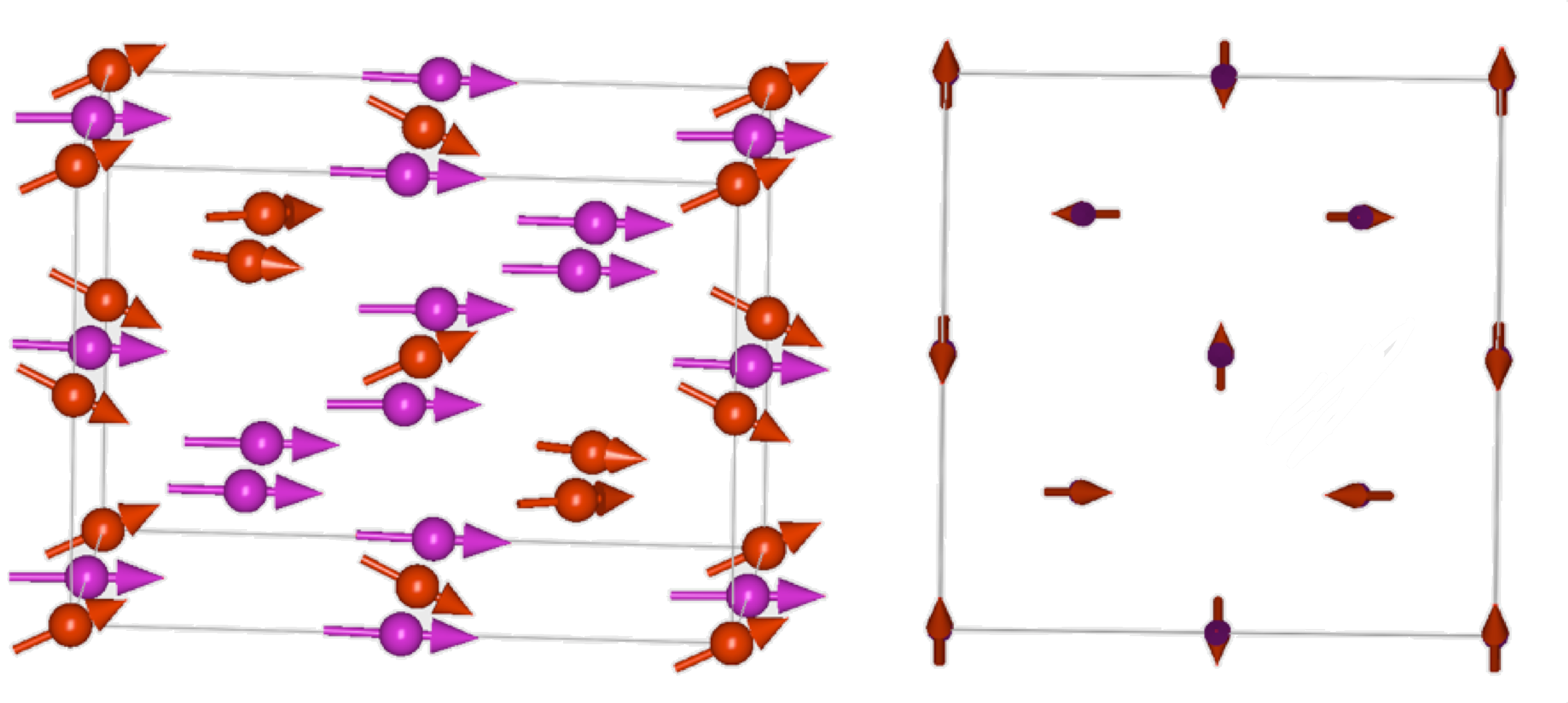}
\end{center}

\caption{(left) The unit cell of \tmo\ showing the magnetic structure in Shubnikov group $I4_11/am'd'$ (no. 141.557). Tm and Mn atoms are represented by violet and red circles. The structure corresponds to the structure listed  in Table~\ref{tab_str}. (right) Projection of the structure on $ab$-plane. Only one type of atoms (labeled as Mn) shows AFM canting.}
\label{mstr_sh}
\end{figure}

The most symmetric solution is given by the maximal tetragonal subgroup $I4_1/am'd'$  that forces all spins of the same type of atom (Mn or Tm) to be the same by symmetry, being the highest allowable by the symmetry solution. Since $\Gamma_4^{+}$ enters the magnetic decomposition twice there are two parameters to be refined: the FM component along z and AFM component in the perpendicular plane for each type of magnetic atoms Tm and Mn. The magnetic structure details and the refined magnetic moments are shown in Table~\ref{tab_str}. As mentioned above we cannot distinguish between Tm and Mn positions in the magnetic structure because they are the only atoms with the magnetic contribution. So, the assignment of the moments per atom type is conditional. Interestingly that the AFM component of moment is converged to the non-zero value for only one type of atom, which has smaller value of FM component and smaller overall spin (it is labeled as Mn in the Table). The only way to distinguish Tm from Mn would be by using the small differences in $Q$ dependence of the magnetic form-factors $f(Q)$ of $\rm Mn^{4+}$ and $\rm Tm^{3+}$ ions, but our experimental accuracy does not allow this. Figure~\ref{mstr_sh} shows the best fit magnetic configuration with the parameters from  Table~\ref{tab_str}.

\begin{figure}
\begin{center}
\includegraphics[width=\figsiz]{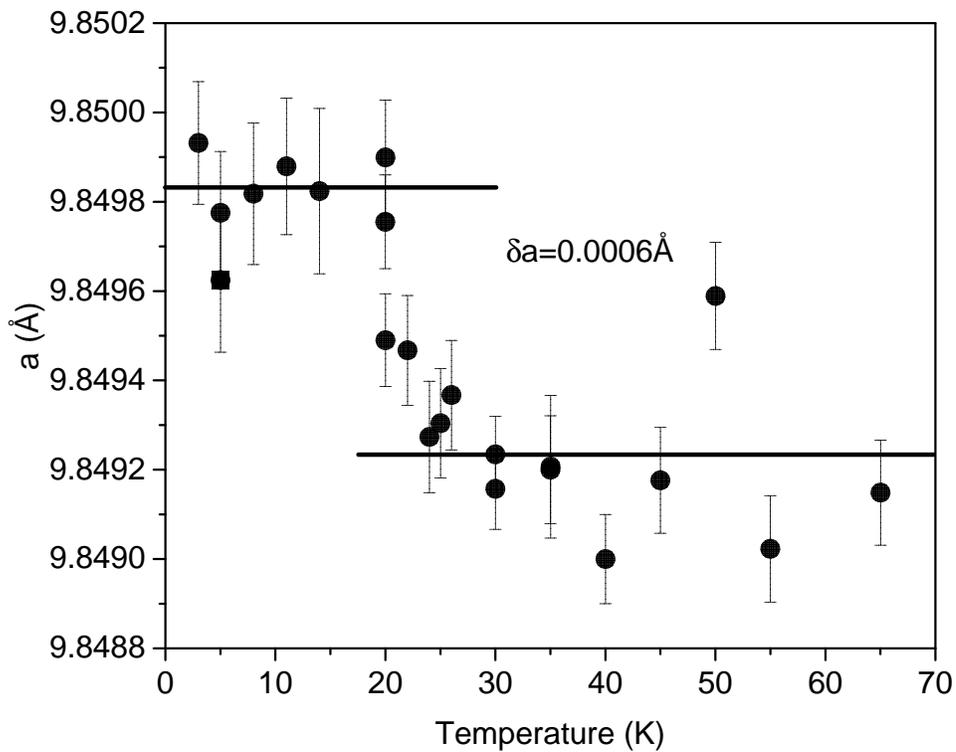}
\end{center}

\caption{Temperature dependence of the lattice constant refined in the cubic metric. }
\label{mtdep}
\end{figure}

\begin{figure}
\begin{center}
\includegraphics[width=10cm]{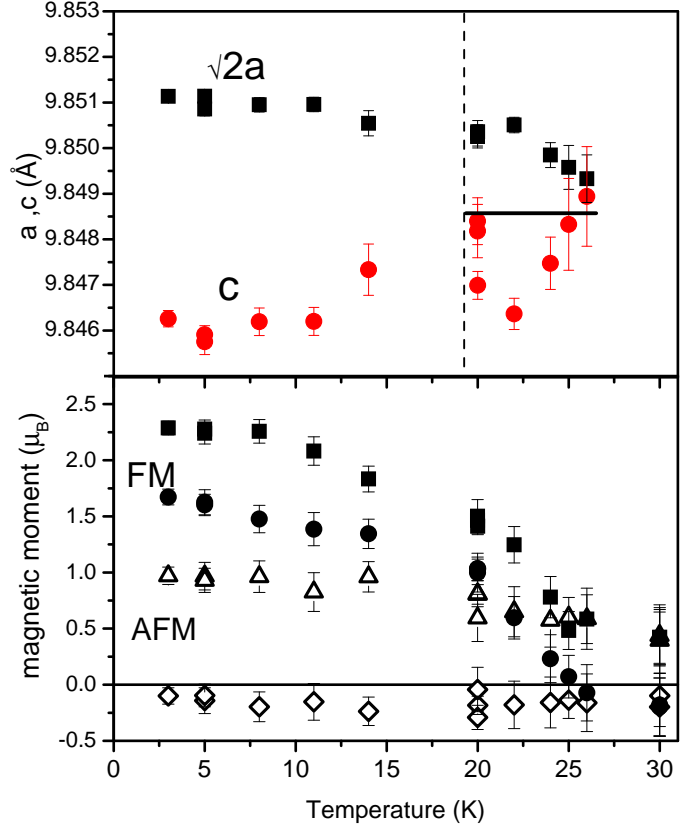}
\end{center}

\caption{(top) Temperature dependence of the lattice constants refined in the tetragonal metric. There is no difference in the fit quality ($\chi^2$) with cubic metric for T$>$19K and no convergence above the magnetic transition  (T $>$25K). The horizontal line shows the value of the lattice constant above $\rm T_C$ refined in the cubic metric. (bottom) Magnetic moment components as a function of temperature. Closed symbols show FM-moments for Tm (squares) and Mn (circles) along $\emph{c}$-direction, open symbols are AFM-moments for  Mn (triangles) and Tm (rhombus) in ($\emph{ab}$)-plane. The assignment of the moments per atom type is conditional. We cannot experimentally distinguish Tm- and Mn-magnetic moments, because they occupy identical symmetry positions.}
\label{tetradist}
\end{figure}

Figure~\ref{mtdep} shows temperature dependence the lattice constant refined in the cubic metric of $Fd\bar{3}m$ space group. One can see an abrupt increase in the lattice constant below $T_C$ implying a presence of a spin-lattice interaction. Ferromagnetic ordering is not possible in the  cubic symmetry and the best magnetic structure solution that we have found is in the tetragonal symmetry. To further check possible tetragonal distortions we performed the following analisys. We have made the fit of the crystal metric as a function of temperature in $I4_1/am'd'$ magnetic group keeping all structure parameters fixed by the cubic structure and with the peak shape parameters fixed by their values above $T_C$. Figure~\ref{tetradist} shows $a$ and $c$ lattice constants (top) and the magnetic moment components (bottom) as a function of temperature. One can see, that below $\rm T_{C}$ there is a pronounced split of the lattice constants. The distortions are significantly bigger than the change in a cubic constant a, because the tetragonal lattice constants a and c increase  and decrease respectively below $\rm T_{C}$, which is accompanied with the increase in the magnetic moments.

\section{Conclusions} In this work we show a novel way of conversion of a thermodynamically stable a hexagonal $\rm TmMnO_{3}$ (space group $P6_{3}cm$) compound  to the metastable $\rm Tm_{2}Mn_{2}O_{7}$ with the cubic pyrochlore-type of structure at 1100 $^\circ$C and 1300 bar oxygen pressure. The unique high temperature high pressure apparatus used in this work allows to synthesize materials in cm$^3$ amount, sufficient to collect good quality neutron scattering data. New studies of magnetic properties of  $\rm Tm_{2}Mn_{2}O_{7}$ by means of a high resolution neutron powder diffraction have shown, that below 25K there is a transition to a ferromagnetically ordered phase with additional antiferromagnetic canting suggesting the crystal symmetry is lower than cubic one. Using symmetry analysis presented here in details, we have found the most symmetry restrictive maximal Shubnikov subgroup $I4_1/am'd'$, that fits the experimental data very well. The magnetic transition is accompanied by well visible magnetostriction effect.

\section*{A{\lowercase{cknowledgements}}}

We acknowledge the allocation of the  beam time at the HRPT
diffractometer of the Laboratory for Neutron Scattering and Imaging (PSI, Switzerland).  The authors thank SNF Sinergia project "Mott physics beyond Heisenberg model" for the
support of this study. The work was partially performed at the
neutron spallation source SINQ.


\begin{mcitethebibliography}{17}
\providecommand*\natexlab[1]{#1}
\providecommand*\mciteSetBstSublistMode[1]{}
\providecommand*\mciteSetBstMaxWidthForm[2]{}
\providecommand*\mciteBstWouldAddEndPuncttrue
  {\def\EndOfBibitem{\unskip.}}
\providecommand*\mciteBstWouldAddEndPunctfalse
  {\let\EndOfBibitem\relax}
\providecommand*\mciteSetBstMidEndSepPunct[3]{}
\providecommand*\mciteSetBstSublistLabelBeginEnd[3]{}
\providecommand*\EndOfBibitem{}
\mciteSetBstSublistMode{f}
\mciteSetBstMaxWidthForm{subitem}{(\alph{mcitesubitemcount})}
\mciteSetBstSublistLabelBeginEnd
  {\mcitemaxwidthsubitemform\space}
  {\relax}
  {\relax}

\bibitem[Subramanian and Sleight(1993)Subramanian, and
  Sleight]{Subramanian1993225}
Subramanian,~M.; Sleight,~A. \emph{Chapter 107 Rare earth pyrochlores, edited
  by Karl A. Gschneidner, Jr. and LeRoy Eyring}; Handbook on the Physics and
  Chemistry of Rare Earths; Elsevier, 1993; Vol.~16; p 225\relax
\mciteBstWouldAddEndPuncttrue
\mciteSetBstMidEndSepPunct{\mcitedefaultmidpunct}
{\mcitedefaultendpunct}{\mcitedefaultseppunct}\relax
\EndOfBibitem
\bibitem[Fujinaka et~al.({1979})Fujinaka, Kinomura, KoizumI, Miyamoto, and
  Kume]{Fujinaka1979}
Fujinaka,~H.; Kinomura,~N.; KoizumI,~M.; Miyamoto,~Y.; Kume,~S. \emph{{Mat.
  Res. Bull.}} \textbf{{1979}}, \emph{{14}}, {1133}\relax
\mciteBstWouldAddEndPuncttrue
\mciteSetBstMidEndSepPunct{\mcitedefaultmidpunct}
{\mcitedefaultendpunct}{\mcitedefaultseppunct}\relax
\EndOfBibitem
\bibitem[Troyanchuk et~al.({1989})Troyanchuk, Derkachenko, and
  Shapovalova]{Troyanchuk1988}
Troyanchuk,~I.; Derkachenko,~V.; Shapovalova,~E. \emph{{Phys. Stat. Sol. A}}
  \textbf{{1989}}, \emph{{113}}, {K249}\relax
\mciteBstWouldAddEndPuncttrue
\mciteSetBstMidEndSepPunct{\mcitedefaultmidpunct}
{\mcitedefaultendpunct}{\mcitedefaultseppunct}\relax
\EndOfBibitem
\bibitem[Subramanian et~al.({1988})Subramanian, Torardi, Johnson, Pannetier,
  and Sleight]{Subramanian1988}
Subramanian,~M.; Torardi,~C.; Johnson,~D.; Pannetier,~J.; Sleight,~A. \emph{{J.
  Solid State Chem.}} \textbf{{1988}}, \emph{{72}}, {24}\relax
\mciteBstWouldAddEndPuncttrue
\mciteSetBstMidEndSepPunct{\mcitedefaultmidpunct}
{\mcitedefaultendpunct}{\mcitedefaultseppunct}\relax
\EndOfBibitem
\bibitem[Shimakawa et~al.({1999})Shimakawa, Kubo, Hamada, Jorgensen, Hu, Short,
  Nohara, and Takagi]{Shimakava1999}
Shimakawa,~Y.; Kubo,~Y.; Hamada,~N.; Jorgensen,~J.; Hu,~Z.; Short,~S.;
  Nohara,~M.; Takagi,~H. \emph{{Phys. Rev. B}} \textbf{{1999}}, \emph{{59}},
  {1249}\relax
\mciteBstWouldAddEndPuncttrue
\mciteSetBstMidEndSepPunct{\mcitedefaultmidpunct}
{\mcitedefaultendpunct}{\mcitedefaultseppunct}\relax
\EndOfBibitem
\bibitem[Gardner et~al.({2010})Gardner, Gingras, and
  Greedan]{Gardner_RevModPhys_2010}
Gardner,~J.~S.; Gingras,~M. J.~P.; Greedan,~J.~E. \emph{{Rev. Mod. Phys.}}
  \textbf{{2010}}, \emph{{82}}, {53}\relax
\mciteBstWouldAddEndPuncttrue
\mciteSetBstMidEndSepPunct{\mcitedefaultmidpunct}
{\mcitedefaultendpunct}{\mcitedefaultseppunct}\relax
\EndOfBibitem
\bibitem[Karpinski({2012})]{Karpinski}
Karpinski,~J. \emph{{Philosophical Magazine}} \textbf{{2012}}, \emph{{92}},
  {2662}\relax
\mciteBstWouldAddEndPuncttrue
\mciteSetBstMidEndSepPunct{\mcitedefaultmidpunct}
{\mcitedefaultendpunct}{\mcitedefaultseppunct}\relax
\EndOfBibitem
\bibitem[Fischer et~al.(2000)Fischer, Frey, Koch, Koennecke, Pomjakushin,
  Schefer, Thut, Schlumpf, Buerge, Greuter, Bondt, and Berruyer]{hrpt}
Fischer,~P.; Frey,~G.; Koch,~M.; Koennecke,~M.; Pomjakushin,~V.; Schefer,~J.;
  Thut,~R.; Schlumpf,~N.; Buerge,~R.; Greuter,~U.; Bondt,~S.; Berruyer,~E.
  \emph{Physica B} \textbf{2000}, \emph{276-278}, 146--147\relax
\mciteBstWouldAddEndPuncttrue
\mciteSetBstMidEndSepPunct{\mcitedefaultmidpunct}
{\mcitedefaultendpunct}{\mcitedefaultseppunct}\relax
\EndOfBibitem
\bibitem[Rodriguez-Carvajal(1993)]{Fullprof}
Rodriguez-Carvajal,~J. \emph{Physica B} \textbf{1993}, \emph{192}, 55;
  www.ill.eu/sites/fullprof/\relax
\mciteBstWouldAddEndPuncttrue
\mciteSetBstMidEndSepPunct{\mcitedefaultmidpunct}
{\mcitedefaultendpunct}{\mcitedefaultseppunct}\relax
\EndOfBibitem
\bibitem[Stokes and Hatch(1988)Stokes, and Hatch]{isot}
Stokes,~H.~T.; Hatch,~D.~M. \emph{Isotropy Subgroups of the 230
  Crystallographic Space Groups}; World Scientific, 1988\relax
\mciteBstWouldAddEndPuncttrue
\mciteSetBstMidEndSepPunct{\mcitedefaultmidpunct}
{\mcitedefaultendpunct}{\mcitedefaultseppunct}\relax
\EndOfBibitem
\bibitem[Campbell et~al.({2006})Campbell, Stokes, Tanner, and Hatch]{isod}
Campbell,~B.~J.; Stokes,~H.~T.; Tanner,~D.~E.; Hatch,~D.~M. \emph{J. Appl.
  Cryst.} \textbf{{2006}}, \emph{{39}}, {607; ISOTROPY Software Suite,
  iso.byu.edu.}\relax
\mciteBstWouldAddEndPunctfalse
\mciteSetBstMidEndSepPunct{\mcitedefaultmidpunct}
{}{\mcitedefaultseppunct}\relax
\EndOfBibitem
\bibitem[Aroyo et~al.(2011)Aroyo, Perez-Mato, Orobengoa, Tasci, De~La~Flor, and
  Kirov]{Bilbao}
Aroyo,~M.; Perez-Mato,~J.; Orobengoa,~D.; Tasci,~E.; De~La~Flor,~G.; Kirov,~A.
  \emph{Bulgarian Chemical Communications} \textbf{2011}, \emph{43}, {183;
  Bilbao Crystallographic Server http://www.cryst.ehu.es }\relax
\mciteBstWouldAddEndPuncttrue
\mciteSetBstMidEndSepPunct{\mcitedefaultmidpunct}
{\mcitedefaultendpunct}{\mcitedefaultseppunct}\relax
\EndOfBibitem
\bibitem[Conder et~al.({2005})Conder, Pomjakushina, Soldatov, and
  Mitberg]{Conder_MatResBul2005}
Conder,~K.; Pomjakushina,~E.; Soldatov,~A.; Mitberg,~E. \emph{{Mat. Res. Bul.}}
  \textbf{{2005}}, \emph{{40}}, {257}\relax
\mciteBstWouldAddEndPuncttrue
\mciteSetBstMidEndSepPunct{\mcitedefaultmidpunct}
{\mcitedefaultendpunct}{\mcitedefaultseppunct}\relax
\EndOfBibitem
\bibitem[Reimers et~al.({1991})Reimers, Greedan, Kremer, Gmelin, and
  Subramanian]{ReimersGreedan_PRB1991}
Reimers,~J.; Greedan,~J.; Kremer,~R.; Gmelin,~E.; Subramanian,~M. \emph{{Phys.
  Rev. B}} \textbf{{1991}}, \emph{{43}}, {3387}\relax
\mciteBstWouldAddEndPuncttrue
\mciteSetBstMidEndSepPunct{\mcitedefaultmidpunct}
{\mcitedefaultendpunct}{\mcitedefaultseppunct}\relax
\EndOfBibitem
\bibitem[Greedan et~al.({1996})Greedan, Raju, Maignan, Simon, Pedersen,
  Niraimathi, Gmelin, and Subramanian]{Greedan_Raju_PRB1996}
Greedan,~J.; Raju,~N.; Maignan,~A.; Simon,~C.; Pedersen,~J.; Niraimathi,~A.;
  Gmelin,~E.; Subramanian,~M. \emph{{Phys. Rev. B}} \textbf{{1996}},
  \emph{{54}}, {7189}\relax
\mciteBstWouldAddEndPuncttrue
\mciteSetBstMidEndSepPunct{\mcitedefaultmidpunct}
{\mcitedefaultendpunct}{\mcitedefaultseppunct}\relax
\EndOfBibitem
\bibitem[Subramanian et~al.({1997})Subramanian, Greedan, Raju, Ramirez, and
  Sleight]{Subra1997}
Subramanian,~M.; Greedan,~J.; Raju,~N.; Ramirez,~A.; Sleight,~A. \emph{{Journal
  De Physique IV}} \textbf{{1997}}, \emph{{7}}, {625}, {7th International
  Conference on Ferrites (ICF 7), Bordeaux, France, Sep 03-06, 1996}\relax
\mciteBstWouldAddEndPuncttrue
\mciteSetBstMidEndSepPunct{\mcitedefaultmidpunct}
{\mcitedefaultendpunct}{\mcitedefaultseppunct}\relax
\EndOfBibitem
\end{mcitethebibliography}
\providecommand*\mcitethebibliography{\thebibliography}
\csname @ifundefined\endcsname{endmcitethebibliography}
  {\let\endmcitethebibliography\endthebibliography}{}

\newpage
\begin{table} \caption{The crystal and magnetic structure parameters in \tmo\ in (a) parent paramagnetic space group $Fd\bar{3}m$ (no. 227, origin choice 2)  at T=30K (b) in magnetically ordered state at T=2~K in Shubnikov magnetic space group $I4_1/am'd'$ (no. 141.557, origin choice 2), which is maximal subgroup of gray parent group for 3D irrep $\Gamma_4^+$ and special direction of order parameter $(a00)$. Atomic displacement parameters $B$ in \AA$^2$ are given for the parent cubic group, magnetic moments $m$ in Bohr magnetons are given for $I4_1/am'd'$ group. Basis transformation from cubic to tetragonal cell reads: (0,1/2,-1/2),(0,1/2,1/2),(1,0,0) with origin shift (-1,1/4,-1/4). Magnetic structure is FM along $z$-axis with AFM canting along $x$ and $y$ axes as shown in Figure~\ref{mstr_sh}. Note, we cannot experimentally distinguish Tm- and Mn-magnetic moments, because they occupy identical symmetry positions. Crystal structure parameters in Shubnikov group are derived from the parent group, according to the above transformation. See the text on the tetragonal distortions. }

\label{tab_str}

\begin{center} \begin{tabular}{l|l|l|l|l} &\multicolumn{2}{c|}{(a) $Fd\bar{3}m$,T=30K}  &\multicolumn{2}{c}{(b) $I4_1/am'd'$,T=5K} \\
 \hline $a$, \AA  &\multicolumn{2}{c|}{9.84959(9)}          &\multicolumn{2}{c}{6.96471}            \\
 $c$, \AA  &\multicolumn{2}{c|}{}                    &\multicolumn{2}{c}{9.84959}            \\

\hline Tm x,y,z              &\multicolumn{1}{l}{16c} &\multicolumn{1}{l|}{0 0 0}             &\multicolumn{1}{l}{8d}            &\multicolumn{1}{l}{0 0 0}              \\
 B,\AA$^2$ m,$\mu_B$         &\multicolumn{2}{l|}{0.094(9)}                                   &\multicolumn{1}{l}{$0,m_y,m_z$}   &\multicolumn{1}{l}{0,-0.05(5),2.31(5)} \\
 \hline Mn x,y,z             &\multicolumn{1}{l}{16d} &\multicolumn{1}{l|}{\ot,\ot,\ot}       &\multicolumn{1}{l}{8c}            &\multicolumn{1}{l}{0,0,\ot}            \\
 B,\AA$^2$ m,$\mu_B$         &\multicolumn{2}{l|}{0.033(23)}                                  &\multicolumn{1}{l}{$0,m_y,m_z$}   &\multicolumn{1}{l}{0,0.81(5),1.66(4)}  \\
 \hline O1 x,y,z             &\multicolumn{1}{l}{8a}  &\multicolumn{1}{l|}{\oe,\oe,\oe}       &\multicolumn{1}{l}{4b}            &\multicolumn{1}{l}{0,\of,$3\over8$}    \\
 B,\AA$^2$ m,$\mu_B$         &\multicolumn{2}{l|}{0.159(24)}                                  &\multicolumn{1}{l}{0,0,$m_z$}     &\multicolumn{1}{l}{0,0,0}              \\
 \hline O2 x,y,z             &\multicolumn{1}{l}{48f} &\multicolumn{1}{l|}{0.4206(6),\oe,\oe} &\multicolumn{1}{l}{8e}            &\multicolumn{1}{l}{0,\of,0.079 }       \\
 B,\AA$^2$ m,$\mu_B$         &\multicolumn{2}{l|}{0.212(10))}                                 &\multicolumn{1}{l}{0,0,$m_z$}     &\multicolumn{1}{l}{0,0,0}              \\
 O2 x,y,z                    &\multicolumn{1}{l}{   } &\multicolumn{1}{l|}{}                  &\multicolumn{1}{l}{16g}           &\multicolumn{1}{l}{$0.2044,x+\of,{3\over8}$} \\
 m,$\mu_B$                   &\multicolumn{2}{l|}{          }                                 &\multicolumn{1}{l}{$m_x,$-$m_x,m_z$}&\multicolumn{1}{l}{0,0,0}\\

\end{tabular} \end{center}
\end{table}

\begin{figure}
\begin{center}
\includegraphics{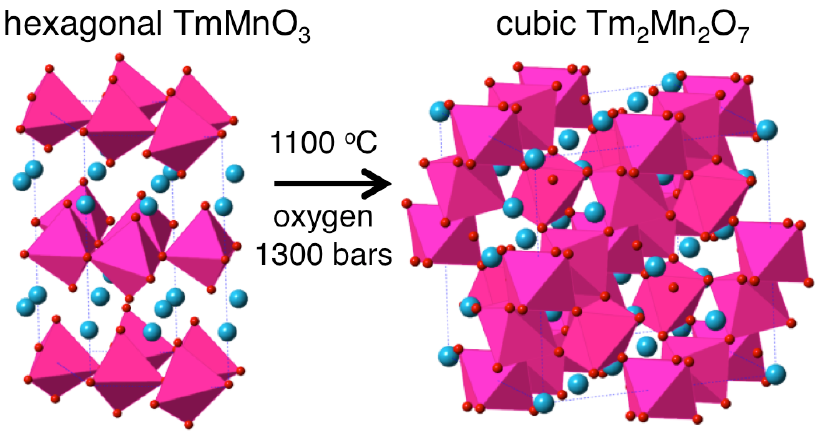}
\end{center}
\caption{For Table of Contents Only}
\label{For Table of Contents Only}
\end{figure}

\end{document}